# Wide-Scale Analysis of Human Functional Transcription Factor Binding Reveals a Strong Bias towards the Transcription Start Site

Yuval Tabach[1,2], Ran Brosh[2], Yossi Buganim[2], Anat Reiner[1], Or Zuk[1], Assif Yitzhaky[1], Mark Koudritsky[1], Varda Rotter[2] and Eytan Domany[1,*]

[1]Departments of Physics of Complex Systems and [2]Molecular Cell Biology, The Weizmann Institute of Science, Rehovot 76100, Israel

[*]Corresponding author.
Email: eytan.domany@weizmann.ac.il    Tel: +972-8-9343964   Fax: +972-8-9344109

**Summary**

**Background:** Elucidating basic principles that underlie regulation of gene expression by transcription factors (TFs) is a central challenge of the post-genomic era. Transcription factors regulate expression by binding to specific DNA sequences; such a binding event is functional when it affects gene expression. Functionality of a binding site is reflected in conservation of the binding sequence during evolution and in over represented binding in gene groups with coherent biological functions. Functionality is governed by several parameters such as the TF-DNA binding strength, distance of the binding site from the transcription start site (TSS), DNA packing, and more. Understanding how these parameters control functionality of different TFs in different biological contexts is an essential step towards identifying functional TF binding sites, a must for understanding regulation of transcription.

**Methodology/Principal Findings:** We introduce a novel method to screen the promoters of a set of genes with shared biological function, against a precompiled library of motifs, and find those motifs which are statistically over-represented in the gene set. The gene sets were obtained from the functional Gene Ontology (GO) classification; for each set and motif we optimized the sequence similarity score threshold, independently for every location window (measured with respect to the TSS), taking into account the location dependent nucleotide heterogeneity along the promoters of the target genes. We performed a high throughput analysis, searching the promoters (from 200bp downstream to 1000bp upstream the TSS), of more than 8000 human and 23,000 mouse genes, for 134 functional Gene Ontology classes and for 412 known DNA motifs. When combined with binding site and location conservation between human and mouse, the method identifies with high probability functional binding sites that regulate groups of biologically related genes. We found many location-sensitive functional binding events and showed that they clustered close to the TSS. Our method and findings were put to several experimental tests.

**Conclusions/Significance:** By allowing a "flexible" threshold and combining our functional class and location specific search method with conservation between human and mouse, we are able to identify reliably functional TF binding sites. This is an essential step towards constructing regulatory networks and elucidating the design principles that govern transcriptional regulation of expression. The promoter region proximal to the TSS appears to be of central importance for regulation of transcription in human and mouse, just as it is in bacteria and yeast.

### INTRODUCTION

Understanding the manner in which transcription is regulated is one of the central challenges of the post-genomic era. Since the most basic regulatory mechanism acts via binding of TFs to the promoter regions of the genes, considerable efforts have been devoted to elucidating TF binding to DNA (Gerland et al. 2002) In spite of very significant advances that were made during the past years, leading to development of novel experimental and theoretical methods to measure and analyze gene expression (Allison et al. 2006) as well as TF binding (see reviews in (Blais and Dynlacht 2005; Ambesi-Impiombato et al. 2006; Elnitski et al. 2006)), several basic questions remain largely unanswered. One of these concerns the extent to which a TF's functionality depends on the location of it's binding site (BS), and another - the relative regulatory importance of different regions of the promoters of higher organisms.

This work makes two distinct but closely related contributions to our understanding of regulation of expression and TF binding. *First*, we study the manner in which the location of a binding site (BS) affects its functionality. One extreme situation is that of a fully location–specific TF, one that affects transcription of its various target genes only if it is bound within a narrow range of distances from the respective TSSs. On the other extreme there may be TFs whose functionality is nearly independent of the BS location, which may vary widely from target to target. Knowing whether a TF belongs to the first or second kind provides important information on the mechanism by which it affects transcription (assignment of a TF to one of these classes may depend on the biological context in which it acts). The first finding we report here is that many TFs are functional when bound within a well defined location window. We also studied the distribution of the BSs of all such location-sensitive TFs, and discovered that the location-specific functional binding sites are much more likely to be observed in the interval (100,-200), i.e. between 100bp downstream to the TSS and about 200bp upstream, than in any window of similar size further upstream (we searched up to 1000bp). The *second* component of the paper is methodological: we present a novel way of searching for functional transcription factor BSs on promoter sequences, in a position-dependent manner. Our method is sensitive enough to reveal the location bias described above. We now proceed to define the concepts used, clarify the question, explain the method and describe how it yields the biological findings.

**Functional binding: biological definition.** A TF may bind to a site on the DNA but this binding event is not necessarily functional. The ideal, biologically sound definition of a functional binding event is that the TF has been shown to bind at the site on a gene's promoter, *and* this binding has been demonstrated experimentally to affect the level of transcription of the gene. Clearly, only functional binding is relevant for understanding regulation of transcription. Such experimental data are, however, scarce and difficult to obtain on a scale that covers all genes and all known transcription factors (and our work poses questions on this scale, as explained below). In principle, had we known all functional BSs, as defined above, for every TF and every gene, we could have provided a definitive answer to the question posed above, regarding the positional distribution of functional BSs. In human the number of known TFs is on the scale of a thousand and the number of genes runs in tens of thousands; hence there are tens of millions of possible TF-promoter pairs. Measuring reliably binding events of all possible TF-promoter pairs is a tall order, but may be forthcoming (Ren et al. 2000) in a few years. However, establishing *functionality* for each bound TF-promoter combination, or even obtaining a large enough unbiased sampling of such pairs, is clearly unrealistic. For this reason we work with a modified "operational" definition of functional binding, that can be used within a computation-based attempt to identify functional BSs.

**Computational approach.** Binding of a TF to DNA at a particular location (BS) is influenced by a variety of factors that affect the energetics of the bound TF-DNA complex. The first factor is the binding sequence – i.e. the sequence of bases that appear at a putative BS. Another factor is the structure (e.g. bending) of the DNA at the BS; obviously epigenetic changes (such as methylation of nucleotides in or near the BS) are very important as well. The proximity of nucleosomes and the methylation, phosphorylation or acethylation states of their constituent histones also affect the chemical environment "seen" by the TF (reviewed in: (Lam et al. 2005) ) The same holds for other proteins that may be bound near the BS, whose presence can either inhibit binding of the TF on which we focus, or enhance binding by modifying it's conformation, or that of the DNA.

Intense recent computational efforts were devoted to identification of BSs for various TFs (Aerts et al. 2003; Elkon et al. 2003; Liu et al. 2003; Ho Sui et al. 2005; Elnitski et al. 2006). These works focused largely on few or a single one of the factors listed above, namely recognizing BSs on the basis of their sequences. These studies relied on information accumulated from various experiments, that reported binding of the TF of interest to DNA segments that have been sequenced (Frech et al. 1997). On the basis of simple statistical analysis of the frequency of appearance of various sequences among observed binding events, a position specific score matrix (PSSM) is constructed for the TF, that allows an easy calculation (see Methods) of a binding score for every DNA sequence. Since sometimes a set of similar sequences is identified as binding a yet unidentified TF, in general each PSSM is associated with a motif *M* (rather than with a TF). A sequence whose score exceeds a threshold is defined as a "hit", e.g. as corresponding to a binding event. Since sequence is only one of the several relevant factors that govern binding, the accuracy of a PSSM-based method to predict TF binding is limited.

Predicting *functional* binding is an even more difficult problem. A TF may bind to a site on the DNA (one that has a high-scoring sequence), without this "stray" binding having any effect on transcription. Since the biological definition of functional binding can be tested only experimentally, for a bioinformatic analysis we need to define functional binding in a way that can be used *in silico*.

**Functional binding: operational definition.** To define functional binding in a computation-based manner we use one or both of two criteria that go beyond the sequence information contained in the PSSM; conservation of the binding motif between species and it's co-occurrence on the promoters of functionally related genes.

*The conservation-based approach* is to compare the regulatory regions of orthologous genes in different species to identify functionally conserved motifs (McGuire et al. 2000; McCue et al. 2001; Rajewsky et al. 2002; Cliften et al. 2003; Kellis et al. 2003; Xie et al. 2005). If a TF is bound at a certain location and the binding is not functional, there is no reason for the binding to be conserved across several species. Using this method, (Xie et al. 2005) showed that BS conserved in four mammalian species exhibited a strong location bias relative to the TSS. However, for a genomic region to be functional, evolutionary conservation is neither necessary (Emberly et al. 2003; Plessy et al. 2005) nor sufficient (Nobrega et al. 2004). Furthermore, their method is restricted to genes that have orthologues over four species, and have TF binding sites with conserved consensus sequences over all these species. For these reasons conservation is a hard test and imposing it is believed to lead to many false negatives (missed functional BSs). For example, their failure to identify the TATA box (see Fig 1D) as a conserved motif at a specific distance from the TSS is indicative of the limitations of relying on conservation alone.

*The other criterion* we use is based on the premise or hypothesis (DeRisi et al. 1997; Hughes et al. 2000) that functionally related genes tend to be co-regulated, and hence if BSs of a TF are highly represented on the promoters of such a group of genes, there is a good chance that this TF is involved in their regulation, and hence these BSs are functional. In order to implement this criterion in a genome-wide study, we have to produce lists of functionally related and co-regulated genes and to test, in a statistically reliable manner, for over-representation of BSs of a TF on their promoters. Such a group of genes may be obtained, for example, from cluster analysis of expression data, which yields groups of co-expressed genes. It is standard procedure (Roth et al. 1998; Zheng et al. 2003; Haverty et al. 2004; Tabach 2005) to perform on such a group of genes a Gene Set Enrichment Analysis (GSEA) (Subramanian et al. 2005) to search for functional Gene Ontology (GO) (Ashburner et al. 2000)classes whose member genes are over-represented in the expression-based gene group. This way one produces a list of genes with the desired property – they are co-expressed and functionally related, and hence are likely to be co-regulated. If a high fraction of these genes' promoters contains BSs of a TF, it is a likely regulator of these genes. The problem with implementing this standard approach in our context is that expression data are available only for a limited number of special conditions, cell types and tissues, and are mostly biased towards several pathologies and diseases. As a result, there are no available experiments and corresponding co-expression lists for all genes and all functional groups. In order to overcome this problem and generate comprehensive and less biased gene lists, *we use the genes that belong to a functional GO class* as our putative co-regulated group $G$. We introduce and implement a novel algorithm that searches for binding sites on the promoters of the genes of such a GO-based group. If the genes that do have a particular binding motif are over-represented in the group, this is viewed (Li et al. 2002) as indicative of functionality of the corresponding BSs.

**Outline of our method and main differences from existing ones.** Existing techniques (Hertz and Stormo 1999; Elkon et al. 2003; Kel et al. 2003; Frith et al. 2004; Haverty et al. 2004) search for BSs by scanning one by one the promoters of various genes and comparing the PSSM-based score of every sequence to a threshold, which is fixed for the entire genome. We follow a different strategy: we execute the search for BSs associated with a motif $M$ *simultaneously* for a group $G$ of (probably) co-regulated genes, but the analysis is carried out *separately* for each location window $W$, obtained by dividing the promoters to short (100, 200 etc bp long) windows. A threshold value $T$ is set, and we count the number of genes in $G$ that scored a hit in $W$ – i.e. a score above $T$ has been recorded, using the PSSM of the tested motif $M$. Next, a hypergeometric test is used to calculate an enrichment p-value, $p(T,G,W,M)$, for the number of hits in $G$, compared to the number of hits expected in $W$, using the same $T$, for a reference group of promoters. The reference group contains all known promoter sequences of the genome. It is very important to use the same window on the promoters of the reference group, since there is a significant systematic variation of the GC content within the region on which we focus (-1000 to +200 bp). For further details, including control of the false discovery rate (FDR), see Methods. Using this method, we performed a comprehensive bioinformatic analysis that encompassed promoters of the 8110 most annotated human genes, 23,400 mouse genes and the known (Quandt et al. 1995) PSSMs of 414 motifs $M$, using all functional GO groups to define the gene sets $G$ that are scanned for enrichment. To reduce false positives further, we repeated the same analysis for the orthologue genes of mouse. Requiring that the same motif is overrepresented in the same location window on the mouse orthologues of the same GO-based genes, is a demanding filter (Xie et al. 2005); hence passing it increases significantly our confidence in the validity of the hit.

Nevertheless, as stated above, the only way to *prove* functionality of a BS is by a series of careful experiments. This, unfortunately, cannot be done in the context of a wide-scale search for binding sites. The best one can do is to present a large number of supporting pieces of evidence and consistency checks; our work is to be viewed in this spirit.

**The biological finding and its relevance.** We found that there are many position-specific TFs, with BSs that satisfy our operational criteria of functionality only within a particular window. The most significant pattern that arises from our analysis is that these location-specific functional BSs are concentrated in a distinct region of about 300bp that includes the TSS; the abundance of such BSs further upstream is significantly lower. The bioinformatic analysis is supplemented and supported by substantive experimental evidence from the literature on a variety of biological systems (Whitfield et al. 2002; Rashi-Elkeles et al. 2006), as well as by a direct experimental test of the dependence of the efficacy of regulation on the distance of the BS from the TSS. There is consensus in prokaryotes on the central regulatory role played by the region close to the TSS, based on evidence from studies in E.Coli (Lodish et al. 2000). As to eukaryotes, arguments supporting the special role of the proximal region have been presented for yeast (Lodish et al. 2000; Harbison et al. 2004). For higher eukaryotes such as human, the difference in terms of functionality between BS found close to the TSS and those found far has still not been fully addressed (Lodish et al. 2000). Several papers showed positional bias towards the TSS by applying the standard approaches; however, most of them dealt with small numbers of *particular* TFs or targets genes (Aerts et al. 2003; Elkon et al. 2003; Liu et al. 2003; Frith et al. 2004; Tabach 2005). As opposed to these studies, several others suggested that remote regions of the mammalian genomes play decisive regulatory roles (Pfeifer et al. 1999; Kimura-Yoshida et al. 2004). Howard et al (Howard and Davidson 2004) suggested that cis-regulatory elements exert their influence over a distance of 100kb; Pfeifer et al (Pfeifer et al. 1999) and Kimura-Yoshida et al (Kimura-Yoshida et al. 2004) argued that the genomic region harboring functionally important regulatory elements can stretch as far as 1Mb in both directions from the TSS, and most recent evidence identified a regulator of transcription that binds to a different chromosome (Ling et al. 2006).

The main novelty of our findings is in their *generality,* as opposed to other works that describe positional bias and specificity of TF binding sites for *particular* TFs or targets genes (Aerts et al. 2003; Elkon et al. 2003; Liu et al. 2003; Frith et al. 2004; Tabach 2005; Carninci et al. 2006). The study of Xie et al (Xie et al. 2005) does make a fairly general statement that complements nicely our findings (see above). Their claims are, however, restricted to genes that have orthologues over four species, and have TF binding sites with conserved consensus sequences over all these species. We claim that positional bias is not limited to strongly conserved sequences; rather, it is a more general and much stronger effect than reported in previous work.

**Overview of the Results.** First, by studying the example of the TATA-box, we demonstrate the difficulties encountered when standard BS search methods are used; difficulties which our approach was designed to overcome. Next we apply the method on more than 8000 human promoters, searching for BSs of all tabulated human TF motifs, in a location and functional GO-group dependent manner. This analysis reveals two distinct classes of GO group – TF motif combinations. The BSs of one of these groups exhibit a strong location bias; analysis of experimental data on cell cycle motifs and their target genes (that belong to this group), confirm our finding of pronounced abundance of the functional BSs in the (0,-200) bp interval. Next we present an experiment that was designed and carried out to test our prediction of location bias. The final result contains re-analysis of another experiment, on NF-κB-regulated genes; here we show how location dependence of functionality explains observed differential expression of genes with similar PSSM-based scores and human-mouse conservation. Furthermore, since new NF-κB targets are also identified, the predictive power of our findings can be put to further experimental test.

**RESULTS**

**Positional bias: the TATA-box as a case study.**

The TATA-box provides a very simple example that demonstrates the advantage of two main components of our method: testing many promoters simultaneously and the need for a position-dependent background. Fig 1 demonstrates several problematic aspects of searching for BSs using standard methods, and it also presents the manner in which our method deals with them, yielding improved robustness and reliability. The PSSM of the TATA motif (see Supplemental Table 1), as provided by MatIspector (Quandt et al. 1995), was used to score each 11bp long window in the interval [+200, -1000] bp with respect to the TSS of a gene. This was done (see Methods) for the promoters of 22,000 human genes, which had been derived from the UCSC genome browser. Scores (see Methods) above -10.4 were identified as hits (i.e. putative binding sites) and marked by a point whose horizontal coordinate represents the location of the hit and the vertical coordinate - the value of its score. The threshold of -10.4 was chosen, as it gave rise to hits in 60% of the promoters, but any higher value of the threshold can be studied. The most striking feature is a high density of hits at about 25 and 35bp upstream from the TSS, which immediately identifies these locations as most likely to be functionally relevant. Clearly, this would not have been evident had we analyzed the promoter of a single gene - which is the strategy used by nearly all BS search methods (Quandt et al. 1995; Kel et al. 2003; Sharan et al. 2004). *Simultaneous viewing of a group of genes* is essential to highlight in a robust way locational bias, which in this case has a very clear meaning: the TATA box is *functional* only when it appears near the TSS. The second feature is the non-uniform spatial distribution of hits; their density becomes significantly higher as the distance from the TSS increases. This is due to the increase of the GC content (in human promoters) from 47% (beyond 1000bp upstream to the TSS) to 55% (at 300bp), up to 64% GC in the region between100bp upstream, to 200 downstream from the TSS (Fig 1B). Hence AT rich motifs tend to be concentrated in the GC poor area, distant from the TSS, and as a result, the number of motifs discovered in the functionally irrelevant ranges, of 100 – 1000bp upstream and TSS to 200bp downstream, may exceed by a factor of 3 to 20 (depending on the threshold used, see Fig 1A and 1C) the proximal functionally relevant motifs. Whereas existing methods use the same threshold for the entire promoter (Hertz and Stormo 1999; Elkon et al. 2003; Kel et al. 2003) and hence generate in this case many false positives, we control this bias by setting *position-* *dependent thresholds*, using as our background model genome sequences from windows of the same location (e.g. TSS to -100bp). Finally, most methods set the threshold value by a uniform criterion for all genes and motifs tested (e.g. set to get a fixed % of hits in the genome (Hertz and Stormo 1999; Elkon et al. 2003; Kel et al. 2003)); obviously, this will generate many false positives for rare motifs and false negatives to very common ones (such as the TATA box). Our method overcomes this: we applied it on the union of two gene sets, which are known to contain proximal TATA motifs. The first set (Amir-Zilberstein et al. 2007) contains 18 genes that appear in Fig 1A and the second (Yang et al. 2007) 66 genes (3 genes appear in both sets). By *determining the threshold separately* for each location (see below), we show that only in the first [0,-100] window we get over-representation of the TATA motif (p-value < $10^{-14}$), with 55 out of the 81 genes passing the threshold. As shown in Fig 1D, if the same value of threshold (-10.27) is used, but in a fixed, location independent manner, in addition to these 55 functional (i.e. proximal) BSs one identifies about 250 stray ones that are not at the functional location. The results obtained by the conservation based method (Xie et al. 2005) are also shown – as expected, it identifies only 12 proximal BSs, and has a relatively large number of false negatives (7 BSs were found out of the window [-100 ,TSS] ).

**Global analysis of Motif, GO and Location Combinations (MGLC).**

Having demonstrated the advantages of our method in a case when the association between location and functionality is very clear, we turn now to present our central results, based on a large-scale analysis. Here we use our operational definition of functionality (see Introduction), to investigate the location bias of the functional BSs associated with 414 vertebrate motifs with known PSSMs (Quandt et al. 1995), in the promoters of 8110 human and 23,400 mouse genes. As described in the Introduction, we used the GO annotation database (Ashburner et al. 2000) to assemble groups of 30 or more unique functionally related genes (smaller groups will not yield good enrichment statistics); there were 134 such different functional GO classes. In total, 7062 human genes (promoters) were assigned to at least one GO group. For each motif *M* and GO group *G*, we performed enrichment analyses (Barash et al. 2001; Subramanian et al. 2005) in 12 consecutive *location-dependent windows W,* each of length 100bp, from 200bp downstream from the TSS to 1000bp upstream. In addition to the 100bp long windows analysed here, we examined also windows of different sizes (of 200bp ,300bp, 400bp and 700bp), with and without overlap (Supplemental Table 2). We found that irrespective of the window size we use, in windows that did not contain the region [100,-200] (or part of it) almost no over represented motifs were found for most of the GO groups. It is important to note that as the window size increases, a larger part of the important functional location range [100, -200] is covered by one window. Because of this reason the number of identified significant BSs increased. On the other hand, using small window gives better location specificity and better control on the GC content of the background – which reduces the number of false positives and yields better location resolution. We decide to focus on the 100bp window but the results for all the windows are presented in the Supplemental Data. Hence 414 ×134 × 12 = 665,712 separate tests were carried out, one for every *(M,G,W)* combination, represented as a path through the top three layers of boxes in Fig 2A. The score *threshold was determined independently* for every *(M,G,W)* combination, as described in Methods and in Fig. 2B. Controlling the False Discovery Rate (Benjamini and Hochberg 1995) (FDR, see Methods) at the level of 0.10, we identified 6331 statistically significant *(M,G,W)* combinations (134 GOs and 412 *M*'s appear in 4130 *M,G* pairs, as listed in Supplemental Table 2). Statistical significance of such an *(M,G,W)* combination means that motif *M* is over-represented in window *W* on the promoters of the genes that belong to the human functional GO group *G*. We refer to such a *significant (M,G,W)* as a Motif, GO-group and Location Combination (MGLC).

Since some of our assumptions may not hold always (e.g. our operational definition of functionality, the assumption that functionally related genes are likely to be co-regulated, etc), a significant fraction of these 6331 MGLCs may be false positives. As explained in the Introduction, *conservation* constitutes an independent and rather strict filter. We submitted our MGLCs to a test of co-appearance for human and mouse (Wasserman et al. 2000; Xie et al. 2005), risking overshoot and producing now a list of MGLCs with many false negatives. By repeating the same analysis for the mouse orthologues of the human genes that had been tested above, we found 785 MGLCs in mouse. The lower number of MGLCs in mouse, compared to human, is probably caused by translation to orthologues; as a result, the GO groups of mouse were of 15% – 30% smaller size. 333 *(M,G)* pairs (see Supplemental Table 2) had an MGLC in human *and* mouse, albeit not necessarily at the same location windows. Each row of Fig 3A corresponds to one of these 333 *(M,G)* pairs, with each bar representing an MGLC. The corresponding 159 motifs (38% of the total 414 motifs in the analysis – see Supplemental Table 2) are associated with 57 GO gene groups (43% of the 134 GO gene groups that were examined). A black bar represents a *conserved* MGLC, one that appears in human *and* mouse, at the same location window; a grey bar – an MGLC in human only *or* mouse only. There are 337 conserved (black) MGLCs, out of the maximal possible number of 785 (p-value< $10^{-320}$); they correspond to 114 different motifs. On the basis of their affiliation with GO groups we assign the MGLCs to one of two classes: the majority belong to a "general GO group" (see below); the distribution of these conserved MGLCs according to location windows (see Fig 3B) reveals a strong bias towards the TSS. The second class of conserved MGLCs are spread somewhat further upstream from the TSS (see Fig 3C). The division of the MGLCs into these two classes is a result of our analysis and is seen clearly in Fig 3A: we noticed that the MGLCs that belong to the second class are based on four functional GO groups, and the *(M,G)* pairs were grouped in Fig 3A accordingly.

The very strong location bias of the MGLCs that belong to the general GO group *is our central finding:* it has been validated (for those genes that have orthologues in human, mouse and rat) using a very different method and a different database of TF binding sites (see Supplemental Material).

**The "General GO Group"; locational bias and experimental verification for cell cycle motifs.** This group contains 177 conserved MGLCs that belong to 93% (53/57) of the significant GO groups: these include cell cycle, immune response, protein biosynthesis as well as others (see Supplemental Table 2). This group shows a strong location bias relative to the TSS: 86% (153/177) of the conserved MGLCs were found in the two promoter windows 0-200 upstream to the TSS, while only 3% (5/177) were found in windows [-200,-1000], and the remaining 19 MGLCs are downstream to the TSS (see Fig 3B). As a further demonstration of our general claim of association between proximal location-specificity of the BS and functionality, we show for the cell cycle related functional class and associated TFs (for which experimental expression data *are* available), that proximal BSs are much more likely to be *functional* than those located further upstream.

We focused on the five cell-cycle associated motifs: CHR, CAAT, and three NFY-family motifs, whose MGLCs were overrepresented close to the TSS in both human and mouse. To test functionality in cell cycle, we used 3,308 genes for which we had both expression data measured during three cell cycles (Whitfield et al. 2002) in HeLa cells, as well as promoter sequences. We calculated the cell cycle periodicity index CCP (Tavazoie et al. 1999) of each of these genes from the expression data, and searched their promoter sequences, from +200 to -1000bp from the TSS, for every one of the 5 motifs. A high CCP indicates strong cell-cycle dependence. We assembled (see Methods) 5×12=60 groups of genes (one for each of 5 motifs *M* studied over 12 non-over lapping windows *W* of 100bp). Each such group had $N_{MW}$ genes on whose promoters we registered a hit (see Methods for details) for the motif *M* in window *W*. Next, for each *MW* combination, the distribution of the CCP scores of the $N_{MW}$ target genes was calculated and compared to the background distribution of the CCP scores over all the 3,308 genes studied. A sequence of these distributions (see Supplemental Figure 1), demonstrates that the distribution of the CCP scores associated with those genes, whose CHR motif is found in the [-100,0] and [0,100] windows, differs significantly from the background distribution (of all the genes), with an enhanced number of genes with CCP scores as high as 5-10. A quantitative measure of the difference between distributions associated with each *MW* versus the background was produced by the Kolmogorov-Smirnov test, yielding a p-value for each *MW*. As seen in Fig 4A, 12 CCP distributions (out of the 60) had p-values below 0.05 and 9 differed from the background at 10% FDR. These include all the 5 motifs that were tested, with windows located near the TSS.

It is gratifying to note the agreement between the statistically significant MW combinations, obtained here from the expression-based CCP distributions, and the results obtained from the sequence-based hyper-geometric over-representation scores (Fig 4B). This agreement demonstrates (a) the reliability of our BS detection method to identify functionally important MGLCs and (b) the functional role of windows that are no more than 300bp upstream of the TSS. These findings concur with our previous observation, that the influence of cell cycle motifs on a cluster of genes obtained from cancer progression is limited to a very specific region, close to the TSS (Tabach 2005).

**The "transcriptional GO group"** contains 160 MGLCs that belong to the following four GO classes: transcription, regulation of transcription, DNA-dependent regulation of transcription and development (Supplemental Table 2). The promoters of the *"transcriptional GO group"* are enriched with motifs that belong to zinc finger proteins (like ZF5, ZF9, ZBP89, ZNF202); with EGR/nerve growth factor induced protein C (such as CKROX, EGR1, EGR3, NGFIC and WT1) and also with some other motifs, like the GC-box motifs NRF, MAZ and MYCMAX. The the transcriptional GO group differs from the general GO group in several aspects. *Firstly*, its motifs are distributed over a much wider region (see Fig 3C), peaking at -300 to -500 bp from the TSS. *Secondly*, the GC content of its motifs (see Methods) is 68% on average, with a minimum of 47%, while the motifs in the general GO group have lower GC content (see the first column of Fig. 3A), with an average of 48% and a minimum of 22% (Fig 3D). Motifs with high GC content are more likely to be found on GC-rich promoters; indeed, the GC content of promoters of the genes that comprise each of the GO classes in the transcriptional GO group is significantly higher (p<0.0001) than expected within a random group of genes (see Methods).We generated two "artificial" PSSMs that contain G/C or C with high probability (Supplemental Table 1) and found that both were over-represented in human and mouse, generating 35 MGLCs in the 4 GO classes that comprise the transcriptional GO group (see Fig. 3A and Supplemental Figure 2). These results suggest that the high GC content in their promoters plays an important regulatory role for genes of the transcriptional GO group. *Thirdly*, for each of the four transcriptional GO classes we found a relatively high number of significantly over-represented TF binding sites. The average of the four numbers was 18, compared to 5 motifs per class in the general GO group. Nearly all these motifs have GC content above 60%.

**Single gene binding sites.** Although our method identifies overrepresentation of a binding motif in a GO-based functional group, we can use the information also for identification of BSs on particular promoters. We prepared a database where for every human gene in our analysis we provide the putative BSs found on it's promoter (including score, location and sequence). In order to get a conservative data base, the putative BSs for each gene were filtered, based on the parameters obtained from the GO groups to which the gene belongs. We register a hit for motif $M$ of a TF on the promoter of a gene if three conditions are satisfied: first, motif $M$ should be significantly over represented in at least one of the GO groups $G$ to which our gene belongs (i.e. there is a corresponding MGLC). Second, the putative BS for motif $M$ that was found on the promoter of our gene should be in the location window of this MGLC, and third; the score of $M$ must pass the corresponding threshold value (that was determined based on the comprehensive GO group analysis). We used 2 different thresholds in this database: 1. 'FDR'- we used the most permissive score threshold that passed FDR<0.1 for the over representation analysis. 2. 'PEAK' – the score threshold that got the best enrichment p-value was used (this score is more conservative). In addition, BSs that belong to MGLCs that are conserved also in mouse where marked as 'motif and location enriched in murine MGLC'. The database can be downloaded from:
http://www.weizmann.ac.il/complex/compphys/downloads/tabach/BS_database.zip

**Direct Experimental verification of our main finding**

This experiment was designed to test our findings and predictions directly. We believe that if functionality of a particular TF's binding is position dependent in a given biological context, its functional BSs will be close to the TSS. Hence we predict that when a functional proximal binding site is moved away from the TSS, it will become less effective.

We designed an experiment to measure transcriptional activity of a TF as a function of the distance between its BS and the TSS. We used the motif $M$ = SRF.01 which binds Myocardin, a transcription co-factor with highly potent ability to activate a set of differentiation-related genes (Wang et al. 2001; Wang et al. 2004). We focused on the target gene SM22, whose expression increased significantly in response to Myocardin (Pipes et al.), and whose transcriptional activation by Myocardin depends on the integrity of specific regulatory motifs called CArG boxes, two of which are present within the SM22 promoter, at distances of ~105bp and ~240bp upstream of the TSS. The proximal motif was shown (Wang et al. 2001) to be significantly more important for the expression of SM22. Here we performed a controlled experiment to eliminate factors such as the possible impact of the large differences between the neighboring regulatory sequences of the two motifs and the (small albeit present) differences between the motif sequences themselves. We generated a set of four plasmids encoding a luciferase reporter gene (*luc*) under the regulation of three tandem repeats of a segment from the SM22 promoter, each of which contained either an intact (*wt*) CArG box, approximately at its center, or a mutated one (*m*). The resulting four constructs thus had identical architectures; while the intact motif is placed at different distances from the TSS (see Fig 5 A).

We transfected HT1080 fibrosarcoma cells that express very low levels of Myocardin, with each of these constructs and measured their transcriptional activity (see Methods). As evident in Fig 5B, co-transfection of Myocardin with the "*m-m-wt-luc*" reporter resulted in dramatic promoter activity increase. In contrast, the Myocardin-dependent activation of "*m-wt-m-luc*" and "*wt-m-m-luc*" constructs that also contain a single CArG box, but at more distal locations, was significantly weaker. Notably, some CArG box-independent activation of transcription exists since even the "*m-m-m-luc*" reporter was activated ~5 fold by Myocardin even though it harbors no functional CArG box. Our results indicate clearly that the transcriptional activity mediated by a regulatory DNA motif is strongly dependent on the distance of this motif from the TSS. The results of the experiment, of decreasing transcriptional effectiveness of the TF with increasing distance of the BS from the TSS, are consistent with our expectations, providing yet one more piece of consistent evidence of their validity.

**Transcriptional efficiency of NF-kB depends on the BS-TSS distance**

Recently Rashi-Elkeles et al (Rashi-Elkeles et al. 2006) identified NF-kB as the most significantly overrepresented motif in the promoters of a cluster of 138 ATM-dependent genes. They focused on 11 out of these 138 genes; on those that have putative NF-kB motifs (associated with the general GO group) in both mouse and in their 11 human orthologues. Ionizing radiation was used to activate ATM, and the measured ATM-induced expression levels of these 11 genes were validated using RT–PCR. All 11 genes belong to the ATM dependent gene cluster reported by (Rashi-Elkeles et al. 2006), and the NF-kB binding motif was identified by PRIMA (Elkon et al. 2003) as a significant hit (Rashi-Elkeles et al. 2006) on all 11 promoters. Nevertheless, the experimental results show a wide variation of the corresponding 11 expression fold changes, ranging from 1.6 to 153 fold induction by ATM activity.

We suggest that BS location and score information distinguishes between genes that are highly induced versus those that have been poorly induced by ATM. Using our algorithm we searched the promoters of all 138 genes for NF-kB binding sites. We found significant over representation of the NF-kB motif in both human and mouse only between the TSS to -200 bp (peaking between -50 and -150, for score threshold between -3 and -5.2; see Fig 6A for the results for mouse). These location bias results fully agree with a specific analysis on NF-κB-regulated immune genes, that examine the positional distribution of NF-κB binding sites (Liu et al. 2003)

In order to examine the influence of both the location and score threshold on the expression of the 138 ATM-dependent genes (Rashi-Elkeles et al. 2006), we examined all their putative NF-kB BSs (see Fig 6B). A group of BSs is clustered at high scores (-3 to -5.2) and in the location window [-200,0] (see box defined by dashed line in Fig 6B). The scores and location windows are almost identical to those found in our over-representation analysis. Inspection of the BSs in this region shows that this group includes those seven genes (out of the 11 marked by solid filled symbols validated by PCR) that had the highest fold induction (Rashi-Elkeles et al. 2006), ranging between 153 and 3.4 , whereas the 4 genes that are outside the box had fold changes between 2.6 and 1.6. With the exception of one gene, ranking the 7 genes in the box by their scores gives the same ordering as the fold induction. The p-value for finding the 7 genes with the highest fold change out of the 11 that were examined in this "location score area" is 0.003. In addition to these 7 genes, we found in the marked location-score box high scoring NF-kB BSs on the promoters of 7 more genes, marked by blank symbols. We predict that these genes are also NF-kB targets and will produce a high fold change. Csf1(Yao et al. 2000) and Nfkbiz (Eto et al. 2003) are two of these 7 genes and, indeed, both are known targets of NF-kB. Three others, Irf1, Ifngr2 and Naf1 are involved in the immune system and hence are likely NF-kB targets. These results demonstrate clearly the importance of the location of the BS, and that our method is able to differentiate between genes that are genuine targets of NF-kB on the one hand and the apparent false positives that were identified by another method.

**SUMMARY AND DISCUSSION**

We introduced a novel method to identify TF binding sites in a location and biological function – dependent manner. Using an operational definition of functional binding sites, we showed that for a large number of TFs, functional binding is location sensitive. These functional binding sites tend to concentrate close to the TSS – within an interval of a few hundred bp. Clearly we cannot rule out that there are functional binding sites also far from this region, but these are most probably not location dependent. These results were demonstrated for both human and mouse.
We found that binding sites that pass the simple test of scoring above a threshold, using the PSSM alone, are much more likely to be conserved across species if they are close to the TSS. They also are more likely to pass both our tests of functionality: over-representation in the promoters of a functional GO class and conservation across species.
A possible point of criticism concerns "discovery bias": the known transcription factor motifs used here represent, probably, only a small fraction of all transcription factors there exist in the human genome. Because of the ways these factors are discovered, these motifs are almost certainly biased towards proteins that bind to proximal promoter regions of genes. However, our results cannot be attributed to this bias. Simply stated, there is no discovery bias that prefers the first 200 nucleotides upstream to the TSS over say the next window of 200-400 bp, or even 400-600; and our main results indicate a very strong bias to the first 200 bp, not the first 2000. Had we claimed overrepresentation of binding sites in the first 2000 bp, discovery bias would have been a valid concern.

Our main finding was obtained by a new motif-discovery method introduced here. The method is characterized by several novel features which make it both flexible and sensitive, identifying hits in a way that takes into account the location, genomic background and a group of potentially co-regulated genes. We performed a high throughput wide scale analysis, for all commonly used motif PSSMs in a large variety of biological functional classes. We found that the GO classes are divided into two groups: the "transcriptional GO group" of 1476 genes, whose promoters have relatively high GC concentrations upstream of 200bp from the TSS (giving rise to a broad distribution of BSs, peaking at ~ -400bp), and a "general GO group", of 6646 genes (with BSs concentrated in the first 200bp upstream of the TSS). The information about the different GO groups and the motifs that were found to be over represented in each are listed in supplemental Table 2

These results were substantiated by showing agreement between our sequence-based analysis with an expression based identification of BSs that are functional in cell cycle regulation. Furthermore, we were able to show, using only NF-κB BS locations and scores, that NF-κB BSs tend to lose their regulatory effect when found at a distance beyond several hundred bp from the TSS, and demonstrated that BS location and score are predictive of the expression level of the activated gene. The principal role played by the proximal region to the TSS emerged also from analysis of BSs that were classified as conserved over human, mouse and rat. Finally, we carried out an experiment whose aim was to test our prediction in the most direct way. By measuring the transcriptional activity of the TF Myocardin when its binding motif is placed at several distances from the TSS on the promoter of the SM22 gene, we demonstrated that the effect of the TF on transcription of the target decreases significantly as the distance of its BS from the TSS increases. Whilst this highly controlled experiment cannot be viewed as a proof of our very general claims, it may serve as a negative control: had we found that Myocardin's transcriptional activity is position dependent, but peaks far (say 5 – 600 bp downstream) from the TSS, this would have contradicted our prediction. Several recent diverse studies strongly support our results that the first 200 bp upstream to the TSS are different, especially in terms of functionality; e.g. the conservation of promoters and BS between species (Carninci et al. 2005; Xie et al. 2005); the high percentage of functional promoter polymorphisms found within the first 100bp (Buckland et al. 2005); the central importance of the first 300bp upstream to the TSS for core promoter activity (Cooper et al. 2006); pronounced depletion of repeatable element concentration near the TSS (Carninci et al. 2005; Marino-Ramirez et al. 2005); identification of about 200bp long nucleosome-free regions on either side of the TSS in yeast (Yuan et al. 2005; Segal et al. 2006) and in higher Eukaryotic genomes (Segal et al. 2006).

TFs are proteins that in order to be functional should be in a physical, contact-based interaction with the transcriptional machinery they regulate, which must have components near the TSS. Therefore, the simplest way to produce such physical interactions is to bind at a location near the TSS, and our findings are consistent with this "Occam's razor". This does not imply, however, that *all* functionally important BSs appear near the TSS. Regulatory regions, such as enhancers, that extend many thousands of bp from the TSS may create complicated stable 3D structures that influence expression by physical interaction with the proximal promoter. Our analyses do indicate, however, that the most prevalent transcriptionally functional mechanisms involve binding in the vicinity of the TSS.

**METHODS**

**Identifying promoter sequences.** DNA sequences upstream of human ORFs were downloaded from the GoldenPath server at UCSC http://genome.ucsc.edu/goldenPath/hg16/bigZips/. Putative regulatory regions for the different genes were obtained (+200 to -1000bp with respect to the TSS). We used as our working set of genes the GeneChip Human Genome Focus Array (Affymetrix, Santa Clara, CA, USA) that represents over 8,500 verified human sequences from the NCBI RefSeq database. Focus chip genes are very suitable for our analysis since they are highly annotated, have low redundancy and their putative promoters are assigned (We identified promoters for 8,110 genes out of the 8,500 of the Human genome Focus chip).

**Scoring a Binding Site.** The score $S(M; \{n\})$ of a putative binding sequence of $L$ nucleotides, $(n_1, n_2, \ldots, n_L)$, for a particular motif $M$ is the log-likelihood score calculated from the motif's PSSM. The entries $m_i(n)$ of the PSSM are the probabilities of finding nucleotide $n_i = C, G, A, T$ at site $i$ :

$$S(M; n_1, .., n_L) = \sum_{i=1}^{L} \log(m_i(n_i))$$

**Determining the threshold $T(M,G,W)$**: set some low initial value for $T$ for the threshold, and count $N(M,G,W)$, the number of genes (of the GO group $G$) for which motif $M$ registered a hit (score > $T$) in the window $W$. Now repeat the process for the same motif, window and $T$, but use the promoters of all the 8,110 human genes to count the number of hits in the same window, $N_0(M,W)$. Now we determine (by the hypergeometric test (Barash et al. 2001; MacIsaac and Fraenkel 2006) $p(M,G,W)$, the p-value for over-representation, i.e. the probability of finding more than $N(M,G,W)$ hits for motif $M$ within window $W$ for $|G|$ genes selected at random from the 8,110. Finally, increase $T$ (steps of 0.1) and repeat the procedure; plots of $p(T)$ are shown in Fig 2B for $G$=mitosis GO group and the motif $M$=NFY1, for all 12 windows. For each of the α = 1,2, …665,712 $(M,G,W)$ combinations we choose as the corresponding threshold $T_α$ the value of $T$ which gives the largest enrichment, i.e. maximal -log $p_α(T)$. Evidently, for the window (TSS- 100bp upstream) we get a clear prominent peak, for the next two windows (-100bp- -300bp) the peak is much smaller and broader, and for all the more distal windows it is within the noise. This way we get the optimal thresholds and corresponding p-values, $p_α(T_α)$, as well as the full set of $p_α(T_α)$, for each $M,G,W$. On the average, 500 different values of $T$ were used for each $M$.

**Controlling the False Discovery Rate:** We assembled the full set of values $p_α(T)$ and applied on it the FDR controlling procedure(Benjamini and Hochberg 1995). The test-statistics corresponding to all the tests cannot be assumed independent of each other, due to several aspects: First, there is an overlap in the genes assembling the different GOs such that two different GO could contain almost the same genes. Second, each score S comprises all the binding sites equal to S or better. As a result, the score S contains all binding sites obtained using more conserved scores. Third, there is spatial correlation among neighboring motifs that are tested with regard to the distance from the TSS. Forth, different PSSM could be very similar and by that to assign the same promoter sequence as a "hit". Since all these imply common biological factor affecting the test, these types of dependencies are potential causes of positive correlation between the test-statistics. The procedure we applied has been shown to control the FDR also for positively correlated test-statistics (Benjamini and Yekutieli 2001).

**Assigning genes to an ($M,W$) group to test the CCP score distribution.** A hit was recorded for gene $g$ when the score of $M$ exceeded a threshold. We described above how the threshold $T(M,G,W)$ was derived. For each gene we have to identify the proper $G$ to be used. For genes that do not belong to any cell cycle related GO class we used the $T$ that was calculated for the cell cycle GO class. If $g$ belongs to a single cell-cycle related GO class, we use this as our $G$. For genes that belong to more than one cell-cycle related GO class we use the one that was most enriched in $M$.

**The GC content** *of a motif* was obtained for a PSSM by calculating for each position the probability to get G or C, and averaging it over the L positions of the PSSM.
*For a GO class* of $n$ genes we calculate the percent (P) of G and C along the $n$ promoters. The p-value for being GC rich was obtained by selecting randomly, 10000 times, groups of $n$ genes and calculating the fraction of such groups that had GC percentage higher than P.

**Construction of reporters for Myocardin from SM22 promoter.** SM22-luciferase reporter constructs were generated by cloning a genomic fragment of the SM22 gene spanning from 166 to -57 bp relative to the TSS into a pGL3 luciferase reporter vector (Promega). Briefly, the genomic region was amplified from human genomic DNA using the primers 5' agcatgcagagaatgtctcg 3' and 5' acagacaggatggggcgctg 3' and ligated into pGEM-T easy vector (Promega). Recognition sequences for either BglII, SmaI or NheI restriction enzymes were added to each primer at the 5' end for the following steps. The Myocardin BS, known as CArG box within the cloned SM22 region, was mutated using QuikChange Site–Directed Mutagenesis kit (Stratagene) and the primers 5' ggtgtCctttcccGaaTtCtggagcc 3' and 5' ggctccaGaAttCgggaaagacacc 3' (Capital letters designate mutated base pairs). The SM22 fragments that contained either an intact Myocardin motif or one which was mutated were then cloned in three consecutive steps into pGL3 vector using the BglII, SmaI and NheI restriction enzymes. Four final luciferase reporters were constructed, all of which contained three tandem repeats of the ~220bp of SM22 regulatory region upstream of the luciferase coding sequence, but each harboring an intact or mutated Myocardin motif in a different order or distances from the luciferase TSS.
The transcriptional activity of these constructs was measured in the HT1080 fibrosarcoma cell line, in which Myocardin is not expressed (Milyavsky et al. 2007). The reporters were transiently transfected into these cells either in the absence or the presence of a small amount of a Myocardin expression vector. To control for the transfection efficiency in each sample, the activity of a cotransfected construct coding for the Renilla-luciferase under the regulation of a CMV promoter was measured.

**Transfections and Reporter Assays** HT1080 fibrosarcoma cells were plated at $2\times 10^4$ cells/well in a 24-well plate, 24 hours prior to transfection. Cells were transfected using FuGene transfection reagent (Roche), with 500ng/well of luciferase reporter construct, 5 ng/well of CMV-Renilla expression vector for normalization of transfection efficiency, and either with or without 2.5ng/well pcDNA3 expression plasmid coding for the mouse Myocardin gene (E.Olson). Cell extracts were prepared 24 hours following transfection Luciferase and Renilla activities were determined using commercial reagents and procedures (Promega). All experiments were conducted in triplicate.

## ACKNOWLEGEMENTS

This work was partially supported by grants from the Ridgefield Foundation, the NIH grant #5 P01 CA 65930-06 and by EC FP6 funding. We thank R. Elkon and Y.Groner for most useful comments.

## FIGURE LEGENDS

**Figure 1** The TATA box score and GC content versus location on the promoters
**A.** Each 11bp-long sequence, located in the interval [+200, -1000] bp with respect to the TSS on the promoters of 22,000 human genes, was scored using the PSSM of the TATA box. Scores above -10.4 were identified as hits and marked by a point whose horizontal coordinate represents the location of the hit and the vertical coordinate represents the value of its score. The high density of hits at about 25 and 35bp upstream from the TSS identifies these locations as most likely to be functionally relevant. The 3 colored dotted lines represent 3 different thresholds (T) and the numbers next to each one of them are the % of TATA BS that were found in the location window [-100,TSS].

**B.** The GC content as a function of location in the interval [+200, -1000] bp with respect to the TSS. The vertical axis represents the percentage of promoters of 22,000 human genes that have a G or a C at each location. Pronounced dips are observed at the TATA box and at the TSS.
**C.** Distribution of identified hits obtained for various values of the threshold, marked by red, green and blue horizontal lines on Fig 1A, with the % of hits found within the proximal peak also given there, for each threshold. As the threshold increases, the overall number of hits decreases, but the relative weight of the proximal peak increases from 8% (for threshold $T=-10.6$) to 27% (for $T = -6.5$).
**D.** Our method identified BSs on 55 out of 81 promoters known to have proximal TATA BSs (see text), in the window [TSS, -100], and none in any of the other location windows (red dot). The threshold selected by our method was -10.27; if this threshold is used uniformly in the entire [200,-1000] interval, about 250 hits are found outside the first upstream interval. The black dots denote the number of hits in and outside the first window, with the numbers next to each point indicating the value of the uniform threshold used. The blue point shows the hits found by Xie et al, demonstrating that their conservation-based method is also effective in eliminating false positives, but has a much larger number of false negatives than our method.

**Figure 2.** Our method of search for BSs.

**A.** Schematic method of search: we used groups of human genes that belong to 134 different functional GO classes *G*, to search for each one of 414 motifs *M* in twelve 100bp-long windows *W* in the interval [+200, -1000] bp with respect to the TSS. In total, $414 \times 134 \times 12 = 665{,}712$ independent analyses were carried out, one for every (*M,G,W*) combination, represented as a path through the top three layers of boxes. Each analysis produced the number of genes of *G* for which the score of *M* in the window *W* exceeded a threshold *T*. The extent of over-representation of such hits was assessed by the hypergeometric test, which compared their number with similar hits in a random selection of |*G*| out of all 8,110 human genes used. A variation of the resulting p-values *p (M,G,W)* with the threshold *T* was studied. All resulting p-values were submitted to FDR analysis, and the statistically significant (*M,G,W*) combinations were intersected with those that were found significant also in mouse.
**B.** The dependence of $-\log[\,p\,(M,G,W)]$ on the score threshold *T* is shown for 12 windows, for *G*=Mitosis GO class and *M*= NFY1. For the window [TSS, -100] bp we get a very prominent peak, for the windows in [-100, -300] bp the peak is much smaller and broader, and for the other windows it is within the noise. We find significant enrichment for the [TSS,-100] window, with the optimal threshold derived from the location of the peak.

**Figure 3** Summary of our main results.
**A.** Using FDR of 0.1 we identify the over-representation of motif *M* on promoters of genes of functional GO class *G*, in location window *W*. Each row corresponds to one of 333 (*M,G*) pairs, composed of 57 GO classes and 159 motifs. Each bar represents an MGLC: grey – either human only *or* mouse only, and black - human *and* mouse. The 57 GO clusters divide into two groups, a "general GO group" of 53 GO classes and a "transcriptional GO group" of 4 GO classes. The motifs *M* of the latter are GC-rich – see first column, where orange denotes GC content above 60%, blue – below 30% and white – between both these values.
**B.** Number of conserved MGLCs of the General GO group, in each window. 93% are in the [-200,0] bp range.
**C.** Same for the transcriptional GO group. The distribution is broad and peaks between 300-500bp upstream from the TSS.
**D.** Distributions of BSs according to their GC content, plotted separately for BSs associated with the two GO groups.

**Figure 4** Testing our results for human cell cycle expression data.
**A.** For 60 groups of genes (each group defined by BS found for one of the 5 cell cycle associated motifs in one of the 12 location windows, see text) the distribution of the expression-based CCP scores (Tavazoie et al. 1999) was compared to a reference background distribution. The resulting 60 p-values are plotted for the different motifs versus their location window. All 5 motifs get p-values less than 0.05 and also pass at FDR of 0.10. Two windows, [0, -100] and [-100, -200] bp, are most significantly over-represented.
**B.** Significant sequence-based hyper-geometric over-representation scores for each of the motif-window combinations. A grey bar represents a significant score (MGLC) for human *or* mouse, for at least one of the cell-cycle related GO classes. A black bar stands for an MGLC in *both* human and mouse**.**

**Figure 5** Direct experimental test of the effect of BS distance from the TSS.
**A.** A graphic representation of four constructs encoding a luciferase reporter (*luc*) gene under the regulation of three tandem repeats of a region from the SM22 promoter, each of which contained either an intact (*wt* in yellow ) or a mutated (*m* in red) BS of the Myocardin transcription factor. The intact motif is placed at different distances from the TSS in each construct.
**B.** Fold activation of the luciferase construct, calculated as the ratio of promoter activity in the presence of Myocardin to the promoter activity in the absence of Myocardin. The *"m-m-wt-luc"* reporter was strongly activated by Myocardin. In contrast, the Myocardin-dependent activation of *"m-wt-m-luc"* and *"wt-m-m-luc"* was significantly weaker.

**Figure 6** Testing our method on ATM-induced expression data
**A.** Analysis of *G*=138 ATM-dependent genes(Rashi-Elkeles et al. 2006) for *M*= NF-kB.01 over representation. The dependence of $-\log[\ p\ (M,G,W)]$ on the score threshold *T* is shown for 23 overlapping windows. Three upstream windows, (TSS-100), (50-150) and (100-200) passed the hypergeometric test at an FDR of 10%.
**B.** putative NF-kB.01 BSs of the 138 ATM-dependent genes; score (vertical axis) versus location (horizontal axis). Our algorithm identifies the location and scores that are indicative of functionality of the BS (rectangle enclosed by the red dashed line). Solid colored symbols mark genes whose expression was validated by PCR. The other genes in the box are marked with blank colored symbols, those that are likely targets of NF-kB outside the box – by X. We list next to the gene names the fold change as measured by the microarray and by PCR (in parentheses). The black dots are belong to the rest of the 138 genes,with putative BS outside the area indicative of functionality.

## SUPPLEMENTAL MATERIAL

**Our findings are reproduced using a different method and other databases.**

To check whether our central findings, regarding the significance of the regions near the TSS, depend on the specific method and data used, we turned to validate our results on the basis of single genes that have orthologues in three different organisms (note that Fig 3A was not based on alignment of the BMs in single genes, but rather on overrepresentation on the GO cluster level). We used the UCSC browser (Kent et al. 2002) (http://genome.ucsc.edu/cgi-bin/hgGateway) which identifies *conserved BM*s – binding sequences with high similarity in human, mouse and rat, that scored above some threshold using the PSSMs obtained from the Transfac Matrix Database (v8.3 created by Biobase) (Wingender et al. 2000) and appear in all three organisms at similar locations on the promoters of the orthologous genes. 409 Transfac motifs had at least one conserved binding site in the regions extending to 3000bp upstream from the closest TSS in the human genome. The total number of conserved BMs of these motifs is 39829; of which 11,829 (30%) were found in the first 200bp, compared to approximately 4% found in each of the other 200bp long windows between 400bp to 3000bp upstream (see Supplemental figure 3). Checking separately the distribution of BMs of each conserved motif shows that 353 out of the 409 (86%) are significantly over-represented ($p < 0.05$, using the binomial distribution) in the first 200bp-long window (Supplemental Table 3). Since the PSSM data contain some redundant entries, we repeated the analysis by using a single PSSM from each one of the 274 matrix families. As expected, by using families instead of single PSSM we obtained very similar results

Note that here we introduced an entirely different method of identifying functionally important BMs, using conservation on individual promoters, as given by an independent database (UCSC browser) and even the database used for BM characterization was different (Transfac). Nevertheless, we still find that the most proximal region to the TSS differs significantly from the rest of the promoter; it contains a significantly larger fraction of conserved motifs. This demonstrates the robustness of our findings regarding the special functional role of the proximal region.

**Legends to the Supplemental Figures and Tables**

**Supplemental Figure 1.** Distributions of CCP scores
Distributions of the CCP scores of 12 groups of genes that contain the CHR binding motif in their promoters, at 12 different location windows (of 100bp) with respect to the TSS. The black curve is our background: the CCP score distribution of all the genes in the experiment(Whitfield et al. 2002)**.**

**Supplemental Figure 2.** Over representation of GC and C rich "artificial" motifs
The hits found in various windows for two "artificial" PSSMs that contain G/C or C with high probability (Supplemental Table1) were over represented in human and mouse. MGLC analysis was done for the different windows and for all GO classes (Supplemental Table2) as explained in the text. Thirty five MGLCs were found, all of which are associated with one of the 4 GO classes that comprise the transcriptional GO group.

**Supplemental Figure 3.** Location distribution of conserved BMs, with high similarity in human, mouse and rat**.**
Location bias analysis of conserved BMs, with high similarity in human, mouse and rat, that were downloaded from the UCSC browser(Kent et al. 2002) and are based on PSSMs obtained from the Transfac Matrix Database v8.3 created by Biobase (Wingender et al. 2000) . The motifs were aligned according to their distance from the closest TSS and the histogram of these distances was plotted. The total number of conserved binding sites of these motifs is 39829; 11829 (30%) of these were found in the first 200bp, compared to about 4% found in each of the other 200bp long windows between 400bp to 3000bp upstream.

**Supplemental Table 1.** The PSSM of the TATA box.
PSSMs used in our work: TATA PSSM provided by MatIspector (Quandt et al. 1995), and two artificial motifs that were generated by us: poly C motif and GC rich motif. The rows are the different nucleotides; the columns represent the nucleotides' position $j$ in the motif. Each entry $m_j(n)$ is the probability to find nucleotide $n$ at position $j$.

**Supplemental Table 2.** Full results for over-representation of binding motifs**.**
 Comprehensive study of the appearances of 414 vertebrate BMs with known PSSMs (Quandt et al. 1995) in the promoters of 8110 human and 23,400 mouse genes. We used groups of human genes that belong to 134 different functional GO classes $G$, to search for each one of 414 motifs $M$ in twelve 100bp-long windows $W$ in the interval [+200, -1000] bp with respect to the TSS. We examined windows of different sizes (200bp ,300bp, 400bp and 700bp), with and without overlap between the windows. In this paper we focous mainly on the analysis of the 100bp window since its gave us better location specificity and better control on the background GC content. In total, for the 100bp windows, 414 ×134 × 12 = 665,712 independent analyses were carried out, one for every (*M,G,W*) combination. The analysis was carried out for human and mouse, and the statistically significant (*M,G,W*) combinations are shown. Each row represents an (*M,G*) pair (first two columns) with a statistically significant over-representation score (passed FDR of 0.10 - see text), in at least one location window in Human (grey with the letter H) or Mouse (grey with the letter M) or in both (black box). The BM symbol and its GC content are also given.

**Supplemental Table 3.** Analysis of conserved BM in human mouse and rat.
39829 conserved BM in human mouse and rat from 409 Transfac PSSMs (v8.3 created by Biobase) (Wingender et al. 2000) were downloaded from UCSC browser (Kent et al. 2002). For each motif (first column) the number of BM found in the first 3000bp from the closest downstream TSS is given in the second column, the number of BMs found in the first 200bp upstream of the TSS are presented in the third column and the corresponding percentage and P-value for this over-representation (derived using the binomial distribution) are given in the fourth and fifth columns.

# Figure 1

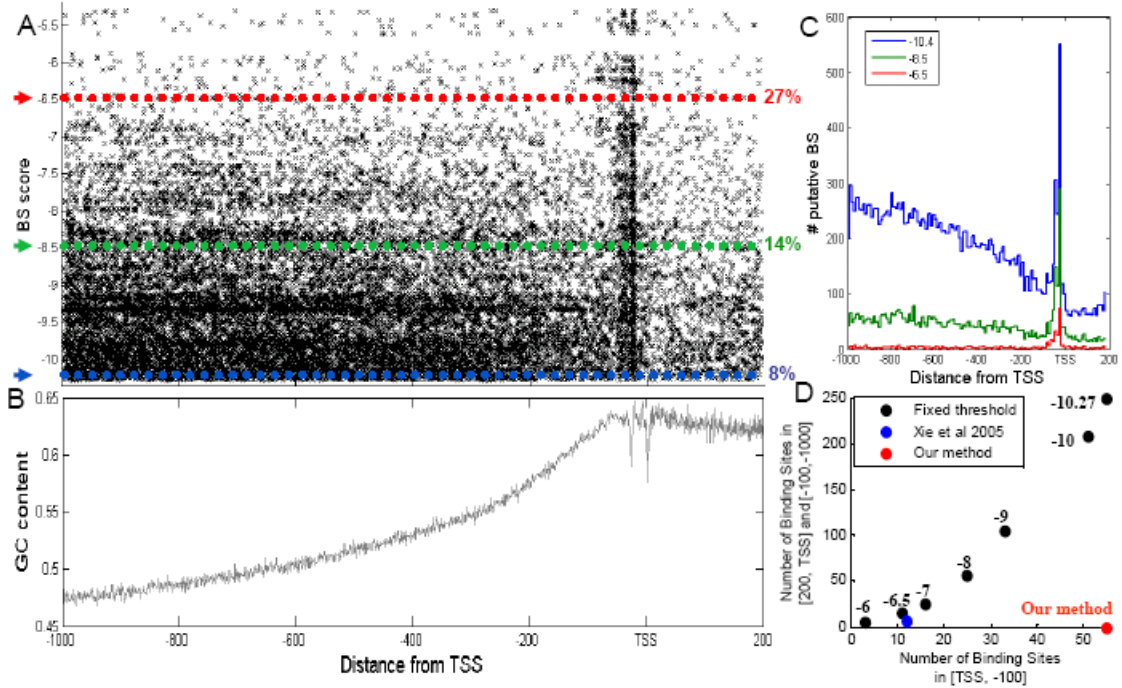

# Figure 2

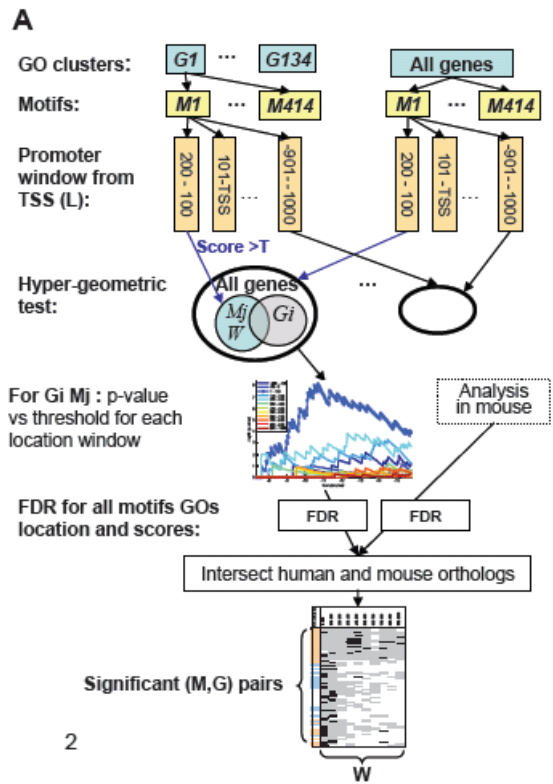

**Figure 2B**

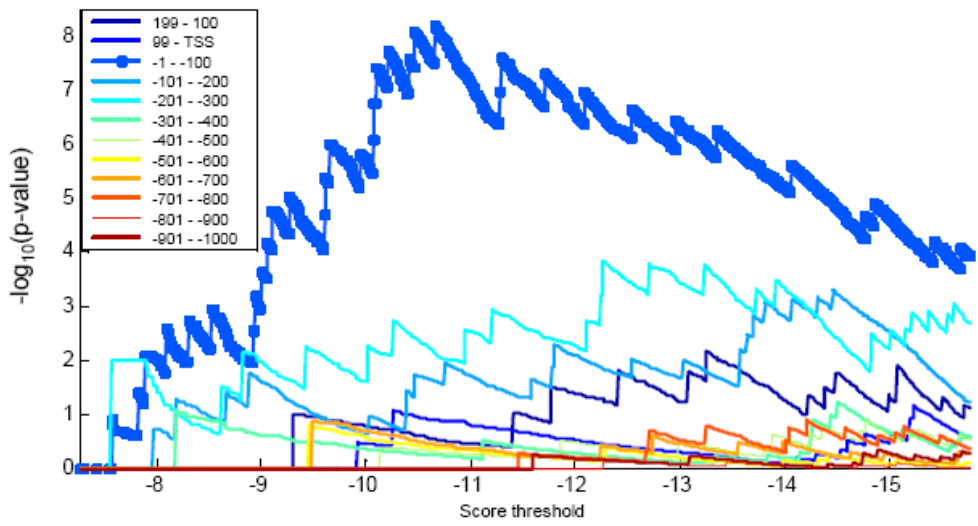

**Figure 3**

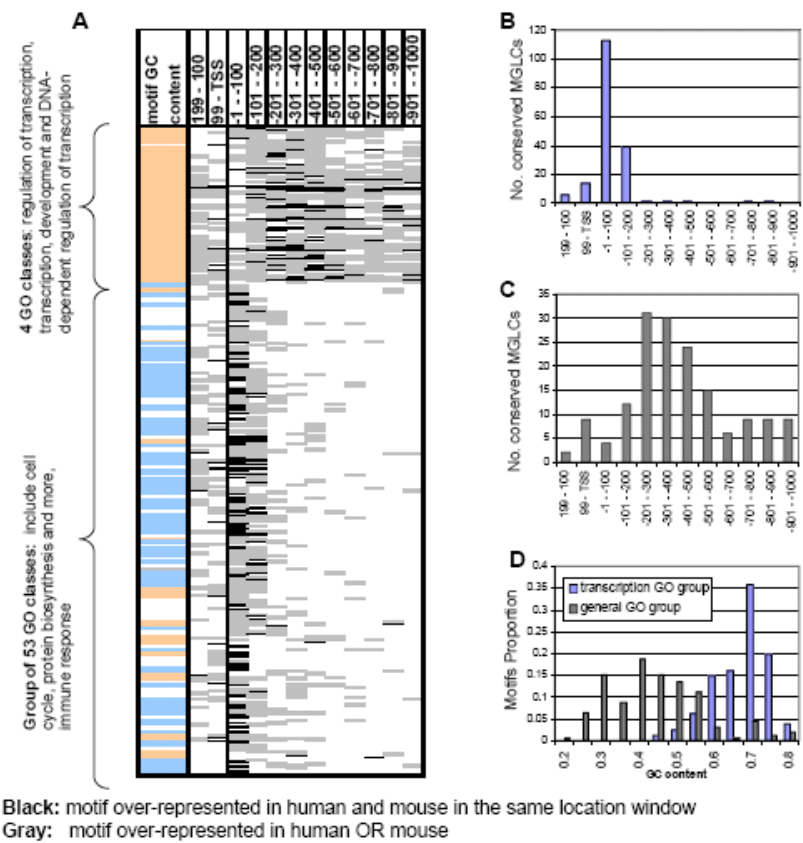

Black: motif over-represented in human and mouse in the same location window
Gray: motif over-represented in human OR mouse

**Figure 4**

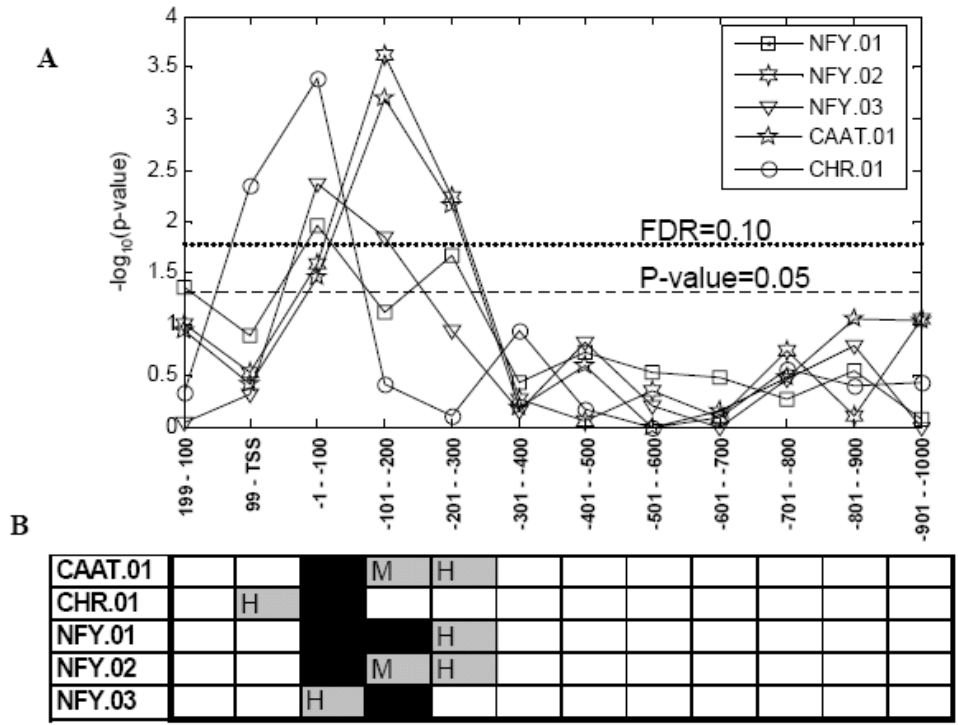

**Figure 5**

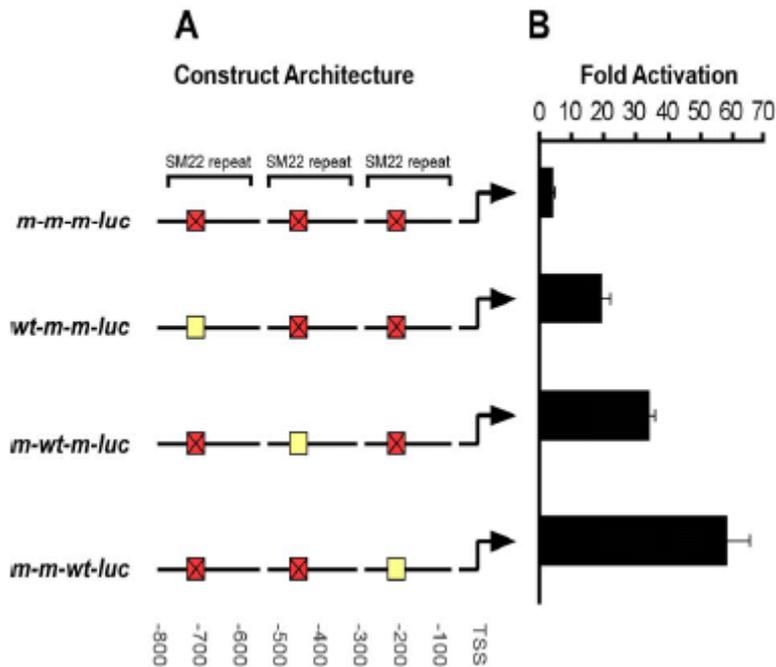

**Figure 6A**

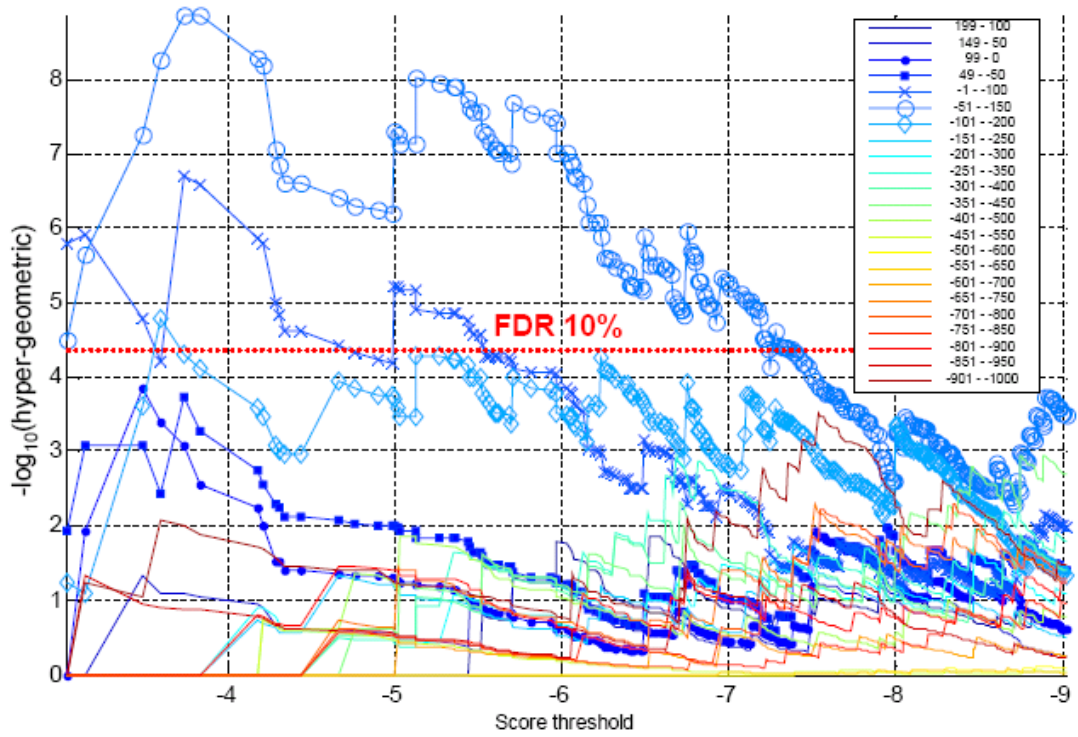

**Figure 6B**

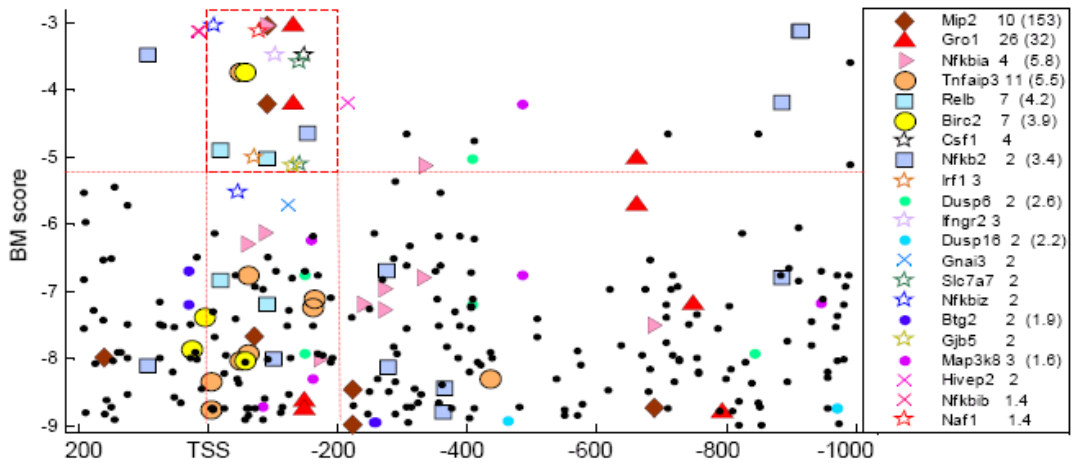

## Supplemental Fig 1

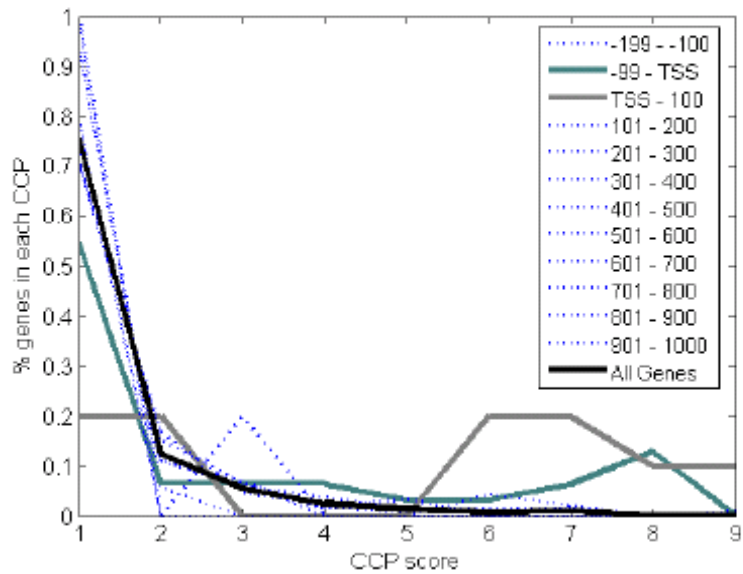

## Supplemental Fig 2

| GO-classes | Motifs | motif GC | -199 - -100 | -99 - TSS | -1 - -100 | -101 - -200 | -201 - -300 | -301 - -400 | -401 - -500 | -501 - -600 | -601 - -700 | -701 - -800 | -801 - -900 | -901 - -1000 |
|---|---|---|---|---|---|---|---|---|---|---|---|---|---|---|
| development (530) | GC_rich | 0.99 | H | 0 | 0 | H | | H | H | M | H | | 0 | H | M |
| regulation_of_transcription (266) | GC_rich | 0.99 | H | 0 | H | | | | H | 0 | 0 | H | | 0 | 0 |
| regulation_of_transcription_DNA_dependent (1066) | GC_rich | 0.99 | H | 0 | 0 | | | | | | H | | | | |
| transcription (930) | GC_rich | 0.99 | H | H | 0 | | | | | | H | | | | |
| development (530) | poly_C | 0.99 | 0 | M | 0 | | H | 0 | H | 0 | 0 | 0 | 0 | M |
| regulation_of_transcription (266) | poly_C | 0.99 | H | 0 | H | | H | H | H | 0 | 0 | 0 | M | 0 |
| regulation_of_transcription_DNA_dependent (1066) | poly_C | 0.99 | H | | H | | | H | | | H | | | H |
| transcription (930) | poly_C | 0.99 | H | | H | H | | H | | | H | | | H |

**Supplemental Fig 3**

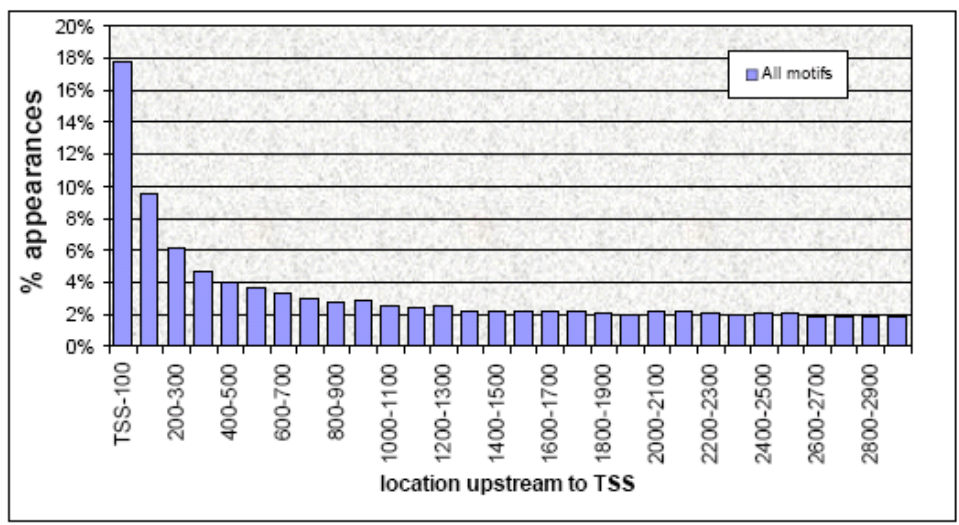